# Title: Command of active matter by topological defects and patterns

**Authors:** Chenhui Peng[†], Taras Turiv[†], Yubing Guo, Qi-Huo Wei and Oleg D. Lavrentovich*

**Affiliation:**

Liquid Crystal Institute and Chemical Physics Interdisciplinary Program, Kent State University, Kent, OH 44242, USA

*olavrent@kent.edu

[†]C.P. and T. T. contribute equally to this work.

**Abstract**: Self-propelled bacteria are marvels of nature with a potential to power dynamic materials and microsystems of the future. The challenge is in commanding their chaotic behavior. By dispersing swimming *Bacillus subtilis* in a liquid-crystalline environment with spatially-varying orientation of the anisotropy axis, we demonstrate control over the distribution of bacteria, geometry and polarity of their trajectories. Bacteria recognize subtle differences in liquid crystal deformations, engaging in bipolar swimming in regions of pure splay and bend but switching to unipolar swimming in mixed splay-bend regions. They differentiate topological defects, heading towards defects of positive topological charge and avoiding negative charges. Sensitivity of bacteria to pre-imposed orientational patterns represents a new facet of the interplay between hydrodynamics and topology of active matter.

**One Sentence Summary:** Patterned orientational order controls concentration, trajectories and flows of swimming bacteria in a liquid crystal.

**Main Text**: Swimming rod-like bacteria such as *Bacillus subtilis* show a remarkable ability to sense and navigate their environment in search of nutrients. They propel in viscous fluids by rotating appendages called flagella, which are comprised of bundles of thin helical filaments. Flagella can also steer the bacterium in a new direction by momentarily untangling the filaments and causing the bacterium to tumble (*1*). Alternating runs and tumbles of bacteria form a random trajectory reminiscent of a Brownian walk. The flows of the surrounding fluid created by bacteria cause their interactions and collective dynamics (*2*). Locally, the bacteria swim parallel to each other but globally this orientational order is unstable, showing seemingly chaotic patterns in both alignment of bacterial bodies and their velocities (*3, 4*). Similar out-of-equilibrium patterns are met in many other systems, universally called "active matter" and defined as collections of interacting self-propelled particles, each converting internally stored or ambient energy into a systematic movement and generating coordinated collective motion (*5-7*). In order to extract useful work from the chaotic dynamics of bacteria (or any other active matter) (*8, 9*), one needs to learn how to control the spatial distribution of particles, and the geometry and polarity of their locomotion.

Placing swimming bacteria in an anisotropic environment of a non-toxic liquid crystal (*10-12*) might be one of the strategies to command active matter. Liquid crystals are anisotropic fluids, with a well-defined nonpolar axis of anisotropy, called the director $\hat{\mathbf{n}} \equiv -\hat{\mathbf{n}}$. A sphere moving in a liquid crystal parallel to the director enjoys the lowest viscous resistance (*13, 14*). It was already demonstrated that the liquid crystal director, either uniform (*10-12*) or spatially distorted (*11, 15*), serves as an "easy swimming" pathway for bacteria. In particular, Zhou et al (*11*) demonstrated curvilinear trajectories of bacteria around obstacles in a liquid crystal, while Mushenheim et al (*15*) rectified the bacterial flows by using nematic cells with the so-called hybrid alignment at the opposite bounding plates. Liquid crystals were also used by Guillamat et al (*16*) as an adjacent layer to an aqueous dispersion of active microtubules to align their active flows. Theoretical models of active systems with orientational order predict that a non-uniform director might cause polar flows (*17, 18*); in particular, Green, Toner and Vitelli (GTV) suggested that unidirectional flows in active nematics can be triggered by mixed splay-bend director deformations.

In this work, we produce spatially varying patterns of a liquid crystal anisotropic environment for swimming bacteria through controlled surface alignment of the director. The patterns are designed with well-defined deformations, either pure bend, or pure splay, or mixed splay-bend. These pre-imposed patterns command the self-propelled bacteria dispersed in such a liquid crystal, in a number of ways, by controlling (i) geometry of trajectories, (ii) polarity of locomotion, and (iii) spatial distribution of bacterial concentration. Bacteria distinguish subtle differences in director deformations that occur over the length scales much larger than their bodies. Namely, their swimming is bipolar in the pure bend and pure splay regions but unipolar in the mixed splay-bend case, matching well the predictions of GTV model (*18*). The bacteria also sense the topological charges of defects in the patterns, moving closer to defects of a positive topological charge and avoiding negative charges.

The patterned anisotropic environment represents thin ($d = 5\,\mu\text{m}$) slabs of a lyotropic chromonic liquid crystal (LCLC) confined between two flat glass plates. The LCLC is an aqueous dispersion of non-toxic disodium chromoglycate (*19*). The bounding plates are pretreated to impose the desired surface alignment of the adjacent LCLC, see Supplement (*20*) and Refs. (*21, 22*). First, they are coated with a layer of photosensitive molecules. This layer is then irradiated with a light beam of linear polarization that changes from point to point, forcing the photosensitive molecules to align in accordance with local polarization. The photoaligned surface molecules align the director of LCLC. The patterns are the same on the top and bottom plates so that the director is two-dimensional (2D).

In a homogeneously aligned liquid crystal, bacteria swim along $\hat{\mathbf{n}}$ in a bipolar fashion, half to the left and half to the right, so that there is no net flow (*11, 12, 15*). In the patterned cells with a pure bend, Fig. 1A, or a pure splay, Fig. 1D, the bacterial locomotion is very similar to the uniform case, being bipolar and parallel to the local $\hat{\mathbf{n}}$, Fig. 1B, 1E, Movies S1-S3. When the number of bacteria in the radial splay configuration is low, they enter and leave the central region freely, Movie S2, but if it is high, they accumulate into a concentrated immobilized disk-like colony, Fig. 1E and Movie S3. In contrast, centers of circular bend remain bacteria-free, Fig. 1B and Movie S1.

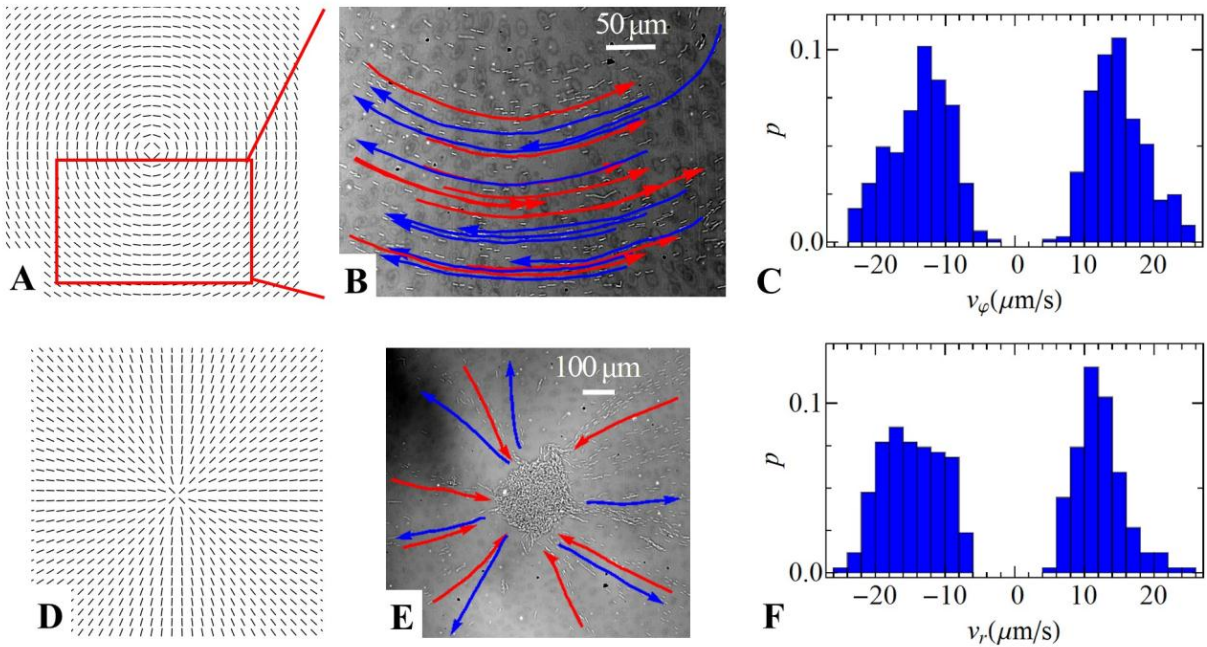

Fig. 1. **Bipolar locomotion of bacteria in patterns of pure bend and pure splay**. (**A**) Director pattern of a pure bend; (**B**) Curcular trajectories of bacteria in bend region; (**C**) Probability distribution of azimuthal velocity of bacteria in the bend pattern; (**D**) Director field of pure splay;

(**E**) Radial trajectories of bacteria moving towards and away from the center; (**F**) Probability distribution of radial velocities of bacteria swimming in the splay pattern.

The bipolar swimming changes to unipolar when the bacteria encounter a pattern in the form of a spiraling vortex with mixed splay and bend, Fig. 2A and Fig. S1. At any point, the local director $\hat{\mathbf{n}}$ makes an angle $45°$ with the radius-vector $\hat{\mathbf{r}}$, spiraling clockwise as one moves away from the center. The vortex is chiral, as it cannot be superimposed on its mirror image. The clockwise spiraling vortex forces counterclockwise circumnavigation of bacteria, producing a net flow, Fig. 2C and Movie S4. The bacteria swim along circular trajectories, i.e. at the angle $45°$ with respect to the surface-imposed $\hat{\mathbf{n}}$, rather than parallel to $\hat{\mathbf{n}}$, Fig. 2B, which is in contrast to the cases of uniform, splayed or bent director fields, Fig. 1.

The spiraling vortex controls also bacterial concentration, by attracting the bacteria to a circular annulus of a finite width. For example, a swarm of 50 bacteria in Fig. 2B occupies a region between $r \approx 15$ and $50 \, \mu m$. The azimuthal velocity $v_\varphi$ within the band is maximum at $r \approx 35 \, \mu m$, Fig. 2C-2D. As the number $N$ of bacteria in the swarm increases (through the influx of bacteria from the periphery) above a threshold of about $N_c = 110$, the circular trajectories start to undulate, Fig. 2G-2H. The undulations are similar to the periodic bend instability described for highly concentrated bacterial dispersions in uniformly aligned LCLC (*11*).

Each spiraling vortex is a point defect of a topological charge $m = 1$ (*23*). The charge $m$ is defined as the number of times the director rotates by $2\pi$ when one circumnavigates the defect (*23*); its sign reflects the direction of rotation with respect to the direction of circumnavigation. Fig. 3A-3B show a periodic pattern of spiraling vortices $m = 1$ and point defects with $m = -1$. The bacteria gather into circulating swarms around $m = 1$ cores, all of which exhibit the same counterclockwise flow, Fig. 3D, 3E and Movie S5.

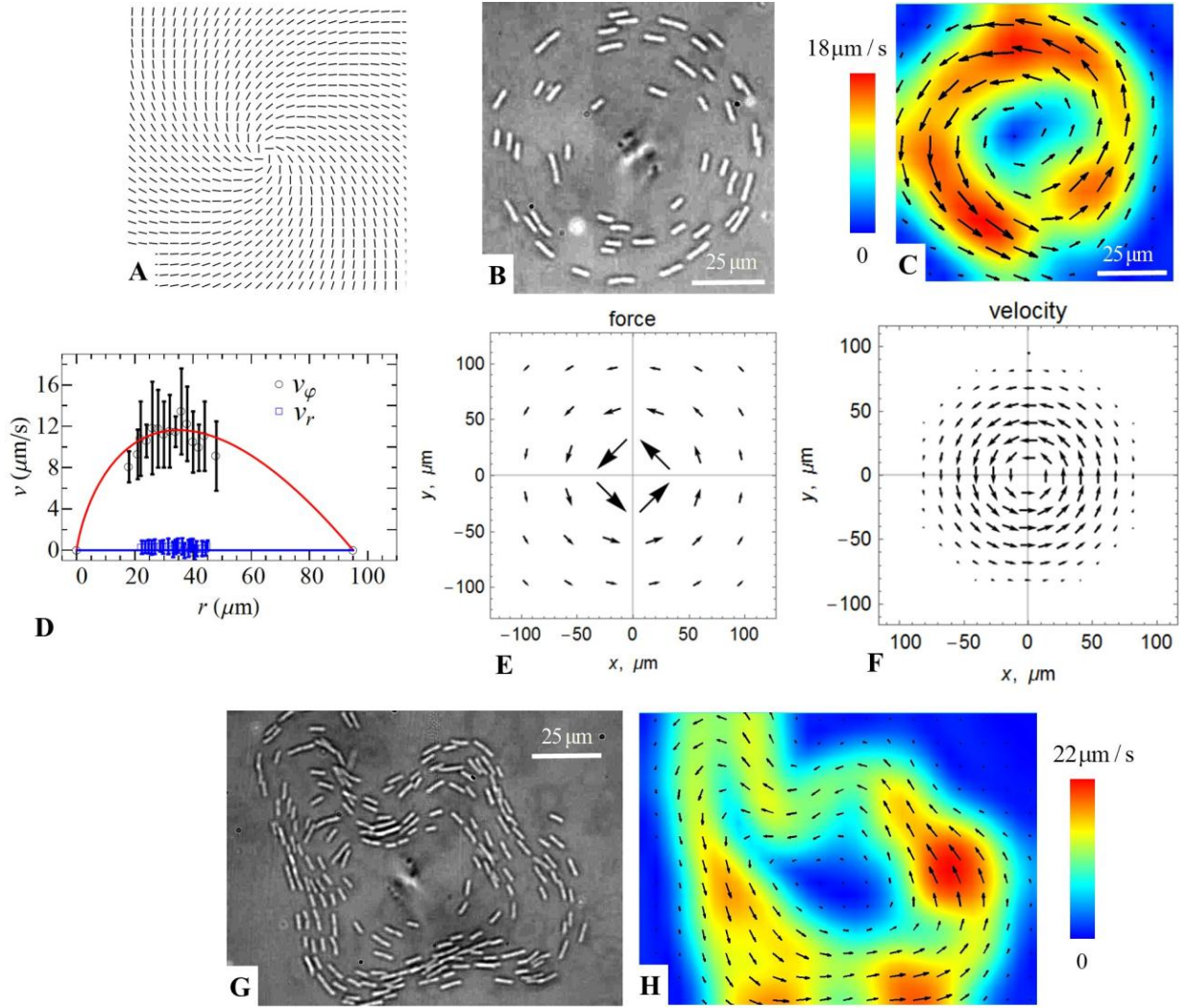

**Fig. 2. Unipolar circular flow of bacteria around a spiraling vortex.** (**A**) Mixed splay-bend director deformation of the vortex; (**B**) circular bacterial swarm enclosing the vortex center; (**C**) map of bacterial velocities; (**D**) radial dependence of the azimuthal and radial velocities of the bacteria; the solid curves are predictions of Eq. (2); (**E**) Active force calculated using Eq. (1); (**F**) Velocity map calculated with Eq. (2). (**G**) Undulation instability of the circular swarm; (**H**) velocity map in the undulating swarm.

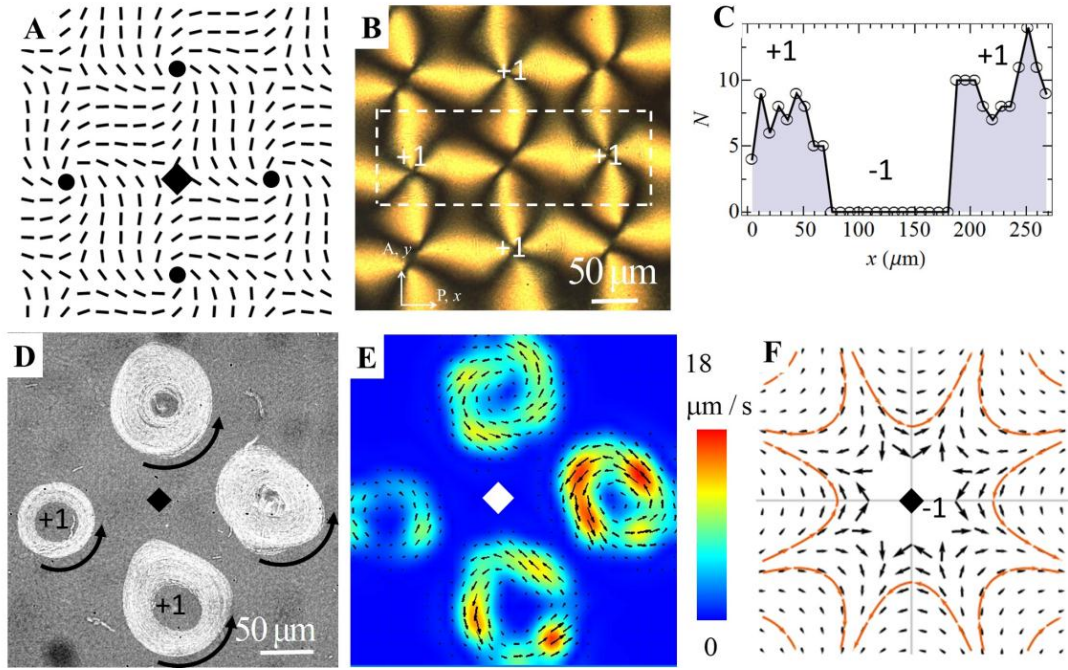

**Fig. 3. Unipolar circular flows of bacteria in the periodic pattern of defects.** (**A**) Director pattern with +1 (cores marked with filled circles) and -1 defects (one core marked by a diamond); (**B**) Polarizing optical microscopy texture of the pattern; (**C**) Spatial modulation of the number of bacteria within the rectangular region in part (B); (**D**) Counterclockwise trajectories of bacteria around four +1 spiraling vortices; (**E**) Map of corresponding velocities; (**F**) Active force calculated with Eq. (1) for a -1 defect.

Unipolar locomotion can also be designed as linear rather than circular. Fig. 4A-4B show a defect pair with $m_1 = 1/2$ and $m_2 = -1/2$. The bacteria prefer to swim from the -1/2 defect towards the +1/2 defect rather than in the opposite direction, Fig. 4C-4D and Movie S6. The +1/2 defect is enriched with the bacteria while the -1/2 defect is deprived of them, Fig. 4E.

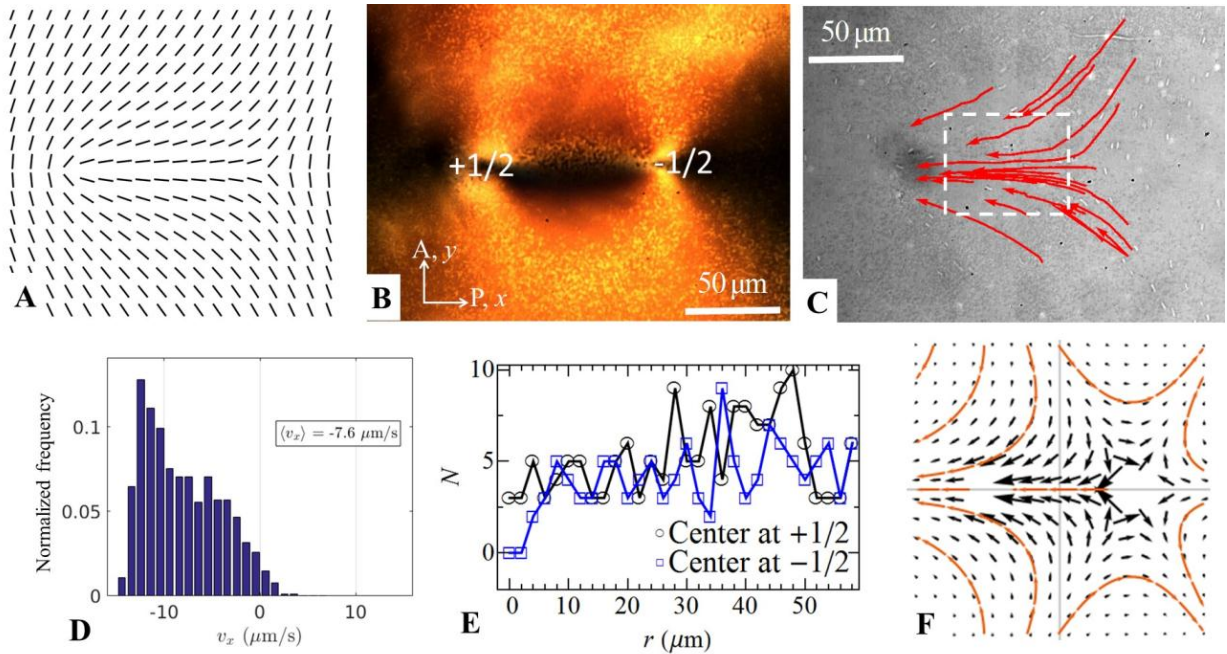

**Fig. 4. Motion of bacteria controlled by a pair of semi-integer defects.** (**A**) Predesigned director field; (**B**) Polarizing microscope texture of the defect pair; (**C**) Trajectories of bacteria moving predominantly from -1/2 defect towards +1/2 defect; (**D**) Polar character of velocities within the rectangular area shown in (**C**); (**E**) Concentration of bacteria vs radial distance from the center of the +1/2 and -1/2 defects; (**F**) Active force calculated for the pair using Eq. (1).

The bacterial dynamics in patterned LCLC shows a remarkable departure from the simple bipolar "swimming along $\hat{\mathbf{n}}$" behavior: (i) bacteria can swim at some angle to the surface-imposed director, as in Fig.2B, 3D; (ii) the locomotion becomes unipolar with the net flows of circular, Fig. 2B, 2C, 3D, 3E and linear, Fig. 4C, 4D, type. Equally important is that (iii) the bacterial concentration varies in space significantly, being high around defects of positive topological charge and low at negative defects, Fig. 3C, 4E.

Emergence of unipolar locomotion in mixed splay-bend regions, Fig. 2-4, is in qualitative agreement with the GTV model (*18*), despite that fact that the model deals with an incompressible system and the experiment shows variation of bacterial concentration across the patterns. A combination of splay $\nabla \cdot \hat{\mathbf{n}}$ and bend $\hat{\mathbf{n}} \times \nabla \times \hat{\mathbf{n}}$ that satisfies the condition $\nabla \times \left[ \hat{\mathbf{n}} \nabla \cdot \hat{\mathbf{n}} - \hat{\mathbf{n}} \times (\nabla \times \hat{\mathbf{n}}) \right] \neq 0$ is predicted to produce a local guiding force (*18*)

$$\mathbf{f} = \alpha \left[ \hat{\mathbf{n}} \nabla \cdot \hat{\mathbf{n}} - \hat{\mathbf{n}} \times (\nabla \times \hat{\mathbf{n}}) \right], \tag{1}$$

where $\alpha = c\sigma$ is the activity, defined by the concentration $c$ of bacteria and the force dipole $\sigma$ exerted by each swimmer on the fluid. For *B. subtilis*, $\sigma$ is negative and estimated to be $\sigma \approx -1\,\text{pN}\,\mu\text{m}$ in an aqueous environment (*24*). Although $c$ varies across the patterns, for a qualitative consideration we assume it constant. Eq. (1) applied to the clockwise spiral vortex in Fig. 2A, $\hat{\mathbf{n}} = \{\cos\theta, \sin\theta\}$, where $\theta = \tan^{-1} y/x - \pi/4$, $x$ and $y$ are Cartesian coordinates, yields $\mathbf{f} = \{0, -\alpha/r\}$ written in polar coordinates $(r, \varphi)$, $r = \sqrt{x^2 + y^2}$, Fig. 2E. The flow velocity $\mathbf{v} = \{v_r, v_\varphi\}$ is obtained by balancing $\mathbf{f}$ with the viscous drag $\eta \nabla^2 \mathbf{v}$:

$$\mathbf{v} = \left\{ 0, \frac{\alpha r}{2\eta} \log\left(\frac{r}{r_0}\right) \right\}, \tag{2}$$

where $r_0$ is the distance at which the velocity vanishes and $\eta$ is the effective viscosity. Equation (2) with $\alpha < 0$ predicts a counterclockwise circulation of bacteria around a clockwise vortex, Fig. 2F, as observed, Figs. 2B, 2C, and 3D, 3E. The velocity field (2) is similar to the velocity around an active gel vortex (*25*) and yields a non-monotonous $v_\varphi(r)$ with a maximum $v_{\max} = -\alpha r_0 / 2e\eta$ at $r = r_0/e$, matching qualitatively the experiment, in which $r_0/e = 35\,\mu\text{m}$ and $v_{\max} \approx 12\,\mu\text{m/s}$, Fig. 2D. The activity/viscosity ratio estimated from Eq. (2) is $\alpha/\eta \approx -0.7\,\text{s}^{-1}$.

The force (Eq. 1) calculated for the pair in Fig. 4F is directed from the -1/2 defect towards the 1/2 defect, reflecting the observed unipolar swimming from -1/2 to +1/2 defect in

Fig. 4C. The force around the -1 defect deflects away from its center, Fig. 3F, again in a qualitative agreement with the observed bacterial avoidance of negative topological charges. For pure splay and pure bend 2D samples, the GTV model (*18*) predicts $\mathbf{f}=0$ and thus no net flows, as in Fig. 1, which shows only bipolar modes of swimming. A more detailed analysis should account for spatial variations of concentration and activity, potential flow-induced realignment of the director, anisotropy and variations of effective viscosities.

The unipolar locomotion of bacteria is observed at relatively low concentration and thus low activity. For example, in the circulating swarm in Fig. 2B, 3D, $c \approx 1.4 \times 10^{15}$ bacteria/m$^3$. This concentration is well below the critical threshold of instabilities described by Zhou et al (*11*) for highly concentrated bacterial dispersions that exhibit undulations and topological turbulence. As the circular swarms attracts more bacteria, increasing $c$ two or three fold, the trajectories start to develop bend undulations, Fig. 2G, signaling that the activity overcomes the stabilizing liquid crystal elasticity (*11*).

To summarize, we demonstrate an approach to control active matter through a patterned anisotropic environment. Self-propelled bacteria sense the imposed patterns of orientational order by adapting their spatial distribution, heading towards topological defects of positive charge and avoiding negative charges, and switching from bipolar locomotion in splayed and bent environments to unipolar locomotion in mixed splay-bend regions. The demonstrated command of active matter by a spatially varying anisotropic environment opens opportunities in designing out-of-equilibrium spatiotemporal behavior, in development of new dynamic materials and systems.

**Acknowledgements**

We are thankful to S. Zhou and G. Cukrov for the help in experiments, I. Aronson and A. Sokolov for illuminating discussions. This work was supported by NSF grants DMR-1507637, DMS-1434185, and CMMI-1436565. C. P., and T. T. performed the experiments. O. D. L. conceived and directed the research. Y. G. and Q.-H. W. provided the metamasks. C.P, T.T., and O.D.L. analyzed the data. All authors participated in discussing and writing the manuscript.